\documentclass[preprint,preprintnumbers,amsmath,amssymb]{revtex4}
\usepackage{graphicx}
\usepackage{dcolumn}
\usepackage{bm}
\begin{document}
\title{Cylindrical solutions in metric f(R) gravity}
\author{ A.~Azadi$^{a}$\footnote{Electronic
address:~amir\_azadi@phymail.ut.ac.ir},
D.~Momeni$^{a}$\footnote{Electronic address:~dmomeni@phymail.ut.ac.ir}
and
M.~Nouri-Zonoz $^{a , b}$ \footnote{ Electronic
address:~nouri@theory.ipm.ac.ir}}
\address{$^{a}$ Department of Physics, University of Tehran, North
Karegar Ave., Tehran 14395-547, Iran. \\
$^{b}$ School of Astronomy and Astrophysics, Institute for Research in
Fundamental Sciences (IPM), P. O. Box 19395-5531 Tehran, Iran.}
\pacs{04.50.+h, 95.36.+x, 98.80.-k}
\begin{abstract}
We study static cylindrically symmetric vacuum solutions in Weyl
coordinates in the context of the metric f(R) theories of gravity.
The set of the modified Einstein equations is reduced to a single
equation and it
is shown how one
can construct exact solutions corresponding to different $f(R)$ models.
In particular the family of solutions with constant Ricci scalar
($R=R_{0}$)
is found explicitly which, as a special case ($R=0$), includes the
exterior spacetime of a cosmic string. Another new solution with constant, 
non-zero Ricci scalar is obtained and its possible relation to the Linet-Tian
solution in general relativity is discussed.
\end{abstract}
\maketitle
\section{Introduction}
As pointed out nicely by Weinberg in his seminal paper \cite{1} on
the cosmological constant problem, "{\it Physics thrives on
crisis}". Perhaps the biggest crisis of the 20-th century physics
which was carried over to the 21-st century is the so called
cosmological constant or dark energy problem: the 120 orders of
magnitude
difference between the observational and theoretical values of the
vacuum energy density. To overcome this crisis, different
proposals have been put forward with the hope of obtaining a
consistent theoretical background to the recent observation of an
expanding universe which is seemingly not accessible through the
standard model of cosmology. Obviously one might trace back this
lack of a proper theoretical explanation to the basic ingredients
of the standard model of cosmology, one of which being the
Einstein field equations derived from the Einstein-Hilbert action.
{\it Modified} or {\it alternative theories of gravity} is the
paradigm under which all those theories which differ from the
Einsteinian gravity are studied. One of these non-Einsteinian
theories, that arose a lot of enthusiasm recently, is the so
called $f(R)$ gravity in which a function, $f(R)$, replaces the
Einstein-Hilbert (gravitational) Lagrangian $R$. It seems that
$f(R)$ actions were first contemplated by Eddington \cite{2} and
later on rigorously studied by Buchdahl \cite{3} in the context of
nonsingular oscillating cosmologies. These theories could be
thought of as a special kind of higher derivative gravitational
theories. Having shown that these models are equivalent to scalar-tensor
models of gravity it is obvious that one should in the first place check their
consistency with the solar system tests of Einstein gravity \cite{solar}. In most of the models
it is not possible to satisfy these tests and at the same time to account for the
accelerated expansion of the Universe without bringing in new degrees of freedom.
Recently it was shown that there are models of $f(R)$ gravity in which one could account
for both the accelerated expansion of the universe and the solar system tests without
introducing the cosmological constant \cite{Hu}-\cite{Sta}.
This is why spherically
symmetric solutions are the most widely studied exact solutions in the
context of $f(R)$ gravity. Apart from this it is interesting, at least from
theoretical point of view, to consider other exact solutions of the modified
Einstein equations of $f(R)$ theory. As in the case of the Einstein-Hilbert action one could
derive field equations in $f(R)$ gravity in two different approaches, the so called {\it metric}
and {\it Palatini} approaches. But in $f(R)$ actions, unlike the Einstein-Hilbert action or its modified version
(one with a  cosmological constant term), the field equations obtained by the two
approaches are not the same in general. In what follows we will be interested only in metric $f(R)$ theories of gravity in which connection is dependent on metric $g_{\mu\nu}$ with respect to which, the action is varied.
In ordinary GR there are not that many exact solutions of the field equations for
a given symmetry. Being higher derivative theory it is not
unexpected to find more exact solutions in $f(R)$ gravity and this
turns out to be  the case for spherically symmetric solutions
\cite{4}. Since cylindrical symmetry is the next symmetry considered
normally in the study of exact solutions in GR (not just for theoretical
reasons but also because they might have physical realization in objects
such as cosmic strings) it seems natural to extend the studies of exact
solutions in f(R) theories in the same direction. Looking for solutions
with a different symmetry, as a
first step, we consider static cylindrically symmetric vacuum
solutions of the $f(R)$ modified Einstein equations in this letter. It is
shown how one can reduce the set of equations into a single
equation which could then be utilized to construct explicit
solutions. For constant curvature solutions, using the general form of a cylindrically
symmetric solution in Weyl coordinates we find, among possible solutions,
a generalized form of a conical (zero curvature) spacetime
as well as two new (non-zero curvature) solutions
with one of their parameters chosen so that it is related naively to the
cosmological constant (section IV). We discuss possible relation to the
so called Linet-Tian (LT) solution \cite{5}-\cite{6} of the modified Einstein field equations
which is the cylindrical analogue of the Schwarzschild-de Sitter spacetime.
\section{Field equations in $f$(R) gravity}
In this section we give a brief review of the field equations in $f(R)$
gravity.
The action for $f(R)$ gravity is given by
\begin{eqnarray}
S=\int(f(R)+\mathfrak{L}_{m})\sqrt{-g}d^{4}x.
\end{eqnarray}
The field
equation resulting from this action in the metric approach, i.e.
assuming that the connection is the Levi-Civita connection and the
 variation is done with respect to the metric $g_{\mu\nu}$, is given by
\begin{eqnarray}
{G}_{\mu\nu}\equiv
{R}_{\mu\nu}-\frac{1}{2}{R}g_{\mu\nu}=\mathrm{T}^{g}_{\mu\nu}
+\frac{\mathrm{T}^{m}_{\mu\nu}}{F(R)},
\end{eqnarray}
where the gravitational stress-energy tensor is
\begin{eqnarray}
\mathrm{T}^{g}_{\mu\nu}=\frac{1}{F(R)}(\frac{1}{2}g_{\mu\nu}(f(R)-R
F(R))+F(R)^{;\alpha\beta}(g_{\alpha\mu}g_{\beta\nu}-g_{\mu\nu}g_{\alpha\beta})),
\end{eqnarray}
with  $F(R)\equiv df(R)/dR$ and $\mathrm{T}^{m}_{\mu\nu}$
the standard matter stress-energy tensor derived from the matter
Lagrangian $\mathfrak{L}_{m}$ in the action (1).
The vacuum equations of motion, i.e. in the absence of matter, are given
by,
\begin{eqnarray}
F(R){R}_{\mu\nu}-\frac{1}{2}f(R)g_{\mu\nu} -
\nabla_{\mu}\nabla_{\nu}F(R)+g_{\mu\nu}\square F(R)=0
\end{eqnarray}
Contraction of the above field equations gives the following relation
between
$f(R)$ and its derivative $F(R)$
\begin{eqnarray}
F(R)R - 2f(R)+ 3 \square F(R)=0,
\end{eqnarray}
which will be employed later both to simplify the field equations and to
find the general
form of the $f(R)$ function.
\section{Cylindrically symmetric vacuum solutions}
Interested in the static cylindrically symmetric solutions of the
vacuum field equations (4), we start with the general form of such a
metric in the
cylindrical Weyl coordinates $(t,r,\varphi,z)$ given by \cite{7};
\begin{eqnarray}\label{metr}
g_{\mu\nu}=diag(-e^{2k(r)-2u(r)},e^{2k(r)-2u(r)},w(r)^{2}e^{-2u(r)},e^{
2u(r)}).
\end{eqnarray}
The corresponding scalar curvature is
\begin{eqnarray}
R=\frac{-w u''+w k'' -w' u' +w'' +w u'^{2}}{w e^{2k}},
\end{eqnarray}
in which $' \equiv\frac{d}{dr}$.
Using equation $(5)$, the modified Einstein equations become
\begin{eqnarray}
F {R}_{\mu\nu}-\nabla_{\mu}\nabla_{\nu}F=\frac{1}{4}g_{\mu\nu}(FR-\square F(R)).
\end{eqnarray}
As in the spherical case, since the metric only depends on the
cylindrical radial
coordinate $r$, one can view equation (8) as a set of
differential equations for functions $F(r)$, $u(r)$, $k(r)$ and $w(r)$.
In this case both
sides are diagonal and hence we have four equations. Differentiating equation
(5) with respect
to $r$  we have the extra, consistency relation for $F(r)$,
\begin{eqnarray}
R {F^\prime}-{R^\prime} F+3(\square{F})^{'}=0.
\end{eqnarray}
Any solution of equation (8) must satisfy this
relation in order to be also a solution of the original modified
Einstein's equations. From equation (8) it is found that
\begin{eqnarray}
\frac{F {R}_{\mu\mu}-\nabla_{\mu}\nabla_{\mu}F}{g_{\mu\mu}}=\frac{1}{4}(FR-\square F(R)).
\end{eqnarray}
In other words the combination $A_{\mu} \equiv \frac{F{R}_{\mu\mu}-\nabla_{\mu}\nabla_{\mu}F}{g_{\mu\mu}}$
(with fix indices) is independent  of the index $\mu$ and therefore
$A_\mu=A_\nu$ for all $\mu,\nu$. This allows us to write the following
independent
field equations;
\begin{eqnarray}
-F^{''}+2F^{'}(k^{'}-u^{'})+F(-\frac{2k^{'}w^{'}}{w}+\frac{w^{''}}{w}+2u^{'2})=0\\
Fw^{2}(-k^{''}-\frac{k^{'}w^{'}}{w}+\frac{w^{''}}{w})+F^{'}(k^{'}w^{2}-ww^{'})=0\\
F
w(-k^{''}+2u^{''}-\frac{k^{'}w^{'}}{w}+\frac{2w^{'}u{'}}{w})+F^{'}(wk^{'}-2u^{'}w)=0,
\end{eqnarray}
corresponding to $A_t = A_r$, $A_t = A_\phi$ and $A_t = A_z$
respectively. Therefore, any set of functions $u(r)$, $k(r)$ and
and $w (r)$ satisfying the above equations would be a solution of
the modified Einstein field equations (8) for a given $F(r)$
satisfying equation (5).
Obviously it is not an easy task to find
a general solution to the above equations, so in the following
section we discuss the simple but important case of  solutions
with constant curvature.
\section{Constant curvature solutions}
It is known that some of the vacuum constant curvature solutions
in $f(R)$ gravity are equivalent to  vacuum solutions in Einstein
theory with the same symmetry. For example it is shown in \cite{4}
that in the spherically symmetric case the corresponding $f(R)$
solutions include the Schwarzschild-de-Sitter space for a specific
choice of one of the constants of integration. For cylindrical
symmetry, in Einstein gravity, static vacuum solutions were found
almost immediately after their spherical counter parts by
Levi-Civita \cite{8} but those with a cosmological constant have
to wait another 60 years to be found independently by Linet \cite{5} and
Tian \cite{6}. Their solution reduces to
that of a cosmic string in the limit $r \rightarrow 0$.\\
Looking for cylindrically symmetric solutions in $f(R)$ gravity,
here we consider the simple but physically important case of
static constant curvature spacetimes. To do so, taking
$R=constant$ in the field equations (11), (12) and (13) , we
arrive at the following set of equations:
\begin{eqnarray}
2u'^{2}+\frac{w''}{w}-2\frac{k'w'}{w}=0\\
k''+\frac{k'w'}{w}-\frac{w''}{w}=0\\
2u''+2\frac{w'u'}{w}-\frac{k'w'}{w}-k''=0.
\end{eqnarray}
From equations (15) and (16) one could obtain the  following two
equations:
\begin{eqnarray}
u'=\frac{1}{2}\frac{w'+c_{2}}{w}\\
k'=\frac{w'+c_{1}}{w},
\end{eqnarray}
in which $c_1$ and $c_2$ are integration constants. By substituting
equations
(17) and (18) in (14) we obtain the following differential equation for
$w(r)$:

\begin{eqnarray}
\frac{1}{2}(\frac{w'+c_{2}}{w})^{2}+\frac{w''}{w}=2\frac{w'(c_{1}+w')}{
w^{2}}.
\end{eqnarray}
One could solve the above equation by inspection and the solutions
discussed below
correspond to solutions with zero and non-zero values of the Ricci
scalar $R$.
\subsection*{Case (1) : solution with $R=0$}
Looking at equation (19)
one could obviously arrange for a solution of $w(r)$ linear in  $r$ (i.e
$w'' = 0$)
in which case the metric functions are given as follows;
\begin{eqnarray}
u=c_{3}\pm\sqrt{\frac{c_{5}}{c_{6}}}\ln(w)\\
k=c_{4}+\frac{c_{5}}{c_{6}}\ln(\frac{w}{c_{6}})\\
w=c_{6}r+c_{7}.
\end{eqnarray}
It could be seen that this is a Ricci flat solution (i.e $R=0$ in
(7)) in which we should identify which one of the constants $c_3$
to $c_7$ correspond to physical parameters of the spacetime and
which ones could be absorbed into the coordinate redefinitions
\footnote{Note that now the constants $c_1$ and $c_2$ could be
written in terms of the constants $c_5$ and $c_6$.}. Substituting
the above functions back into the metric form (\ref{metr}) we
obtain
\begin{eqnarray}\label{metric2}
ds^2=e^{-2(c_3\pm \sqrt{\frac {c_5}{c_6}}ln \rho)}\left(e^{2({\tilde
c_4}+\frac {c_5}{c_6}ln \rho)}(\frac{d{\rho}^2}{c_6}-dt^2)+{\rho}^2 d\phi^2\right) +
e^{2(c_3\pm \sqrt{\frac {c_5}{c_6}}ln \rho)}dz^2,
\end{eqnarray}
in which ${\tilde c_4} = c_4 - \frac{c_5}{c_6} ln c_6$ and  $\rho \equiv
w = c_6 r$ is the
new radial coordinate by setting  $c_7=0$ without any loss of
generality.
One can also show that through the following redefinitions of the
constants and the coordinates

\begin{eqnarray}\label{redef}
m = \sqrt{\frac{c_5}{c_6}}\\
{\tilde t} = e^{{\tilde c_4} - c_3} A ^{\frac{1}{m(m \mp 1) + 1}} t\\
{\tilde z} = e^{c_3} A ^{\frac{\mp m}{m(m \mp 1) + 1}} z\\
{\tilde \phi} = e^{-c_3}A ^{-\frac{1\mp m}{m(m \mp1) + 1}} \phi\\
{\tilde {\rho}} = A ^{\frac{1}{m(m \mp 1) + 1}}\rho\\
A = e^{\frac{{\tilde c_4} - c_3}{c_6}},
\end{eqnarray}

the above metric reduces to

\begin{eqnarray}\label{metric3}
ds^2 = {\tilde {\rho}}^{2m(m \mp 1)} (d{\tilde {\rho}}^2 - {d\tilde t}^2 ) +
{\tilde {\rho}}^{2 \mp 2m}{d{\tilde \phi}}^2 + {\tilde {\rho}}^{ \pm 2m}{d\tilde
z}^2,
\end{eqnarray}

and on applying the following complex transformation \cite{7}
\begin{eqnarray}
\tilde t \longrightarrow i\tilde z \;\;\ \tilde z \longrightarrow
i\tilde t,
\end{eqnarray}
it transforms into the following well known metric
\begin{eqnarray}\label{metric4}
ds^2 = {\tilde {\rho}}^{2m(m \mp 1)} (d{\tilde {\rho}}^2 + {d\tilde z}^2 ) +
{\tilde {\rho}}^{2 \mp 2m}{d{\tilde \phi}}^2 - {\tilde {\rho}}^{ \pm 2m}{d\tilde
t}^2,
\end{eqnarray}
which is formally similar to the Levi-Civita's static cylindrically symmetric
solution in GR normally written without $\pm$
sign but with the constant $m$  taking both positive and negative
values.
It should also be noted that the range of the variable $\tilde \phi$ is not
in general $(0, 2\pi]$,
not even for the flat cases of $m=0, 1$. \\
In the case of  $m=0$  the spacetime is conical with a deficit angle
corresponding to the
exterior metric of a cosmic string  with the following line element
\cite{9}
\begin{eqnarray}\label{metric5}
ds^2 = (d{\tilde {\rho}}^2 + {d\tilde z}^2 ) + {a_0}^2  {\tilde
{\rho}}^2{d{\phi}}^2 - {d\tilde t}^2,
\end{eqnarray}
in which $a_0 = c_6 e^{-c_4}$ is the conical parameter related to
the {\it gravitational mass per unit length of the spacetime, $\eta$}, as
\cite{10}
\begin{eqnarray}
a_0 = c_6 e^{-c_{4}} = 1-4\eta,
\end{eqnarray}
such that  $0 < a_0 < 1$ for $0 < c_{4} < \infty$ (taking $c_6 = 1$).
This metric, exposing the geometry around a straight
cosmic string, is locally identical to that of flat spacetime however it
is not
globally Euclidian
since the angle ${\tilde \phi} = a_0 \phi$ varies in the range $0\leq
{\tilde \phi}< B$
where $B=2\pi a_0 =2\pi(1-4\eta)$.\\
From equation (5) it is seen that this metric is a solution for
any form of $f(R)$ for
which $f(R=0)=0$, in other words, $f(R)$ must be a linear
superposition of $R^{n}$ with $n \geq 1$ of
which the commonly considered model is $f(R) = R + \frac{R^{2}}{\Lambda}$ and
obviously it is not
a solution of the widely studied model for which $f(R)=R - \frac
{{\mu^4}}{R}$.
\subsection*{Case (2) : solutions with $R = constant \neq 0$}
We discuss two solutions of this type here:\\
{\bf A})-The first solution could be obtained by noting the simplifying fact that
equation (16) is satisfied for $k=2u$ so that equations (14) and (15) reduce to,
\begin{eqnarray}
2u'^{2}+\frac{w''}{w}-4\frac{u'w'}{w}=0\\
2u''+\frac{2u'w'}{w}-\frac{w''}{w}=0,
\end{eqnarray}
leading to the following equation,
\begin{eqnarray}\label{n1}
u''+u'^{2}-\frac{u'w'}{w}=0,
\end{eqnarray}
and consequently to the following relation between functions $u$ and $w$,
\begin{eqnarray}\label{n2}
e^u = D \int w d\rho,
\end{eqnarray}
in which $D$ is a constant to be determined later. On the other hand using (\ref{n1}), from equation (7) we have,
\begin{eqnarray}
R \equiv R_0 = \frac{w''}{w}e^{-4u}.
\end{eqnarray}
Substituting from (\ref{n2}) we have;
\begin{eqnarray}\label{n4}
\frac{w''}{w}= R_0 (D \int w d\rho)^4
\end{eqnarray}.
A solution to this integro-differential equation is given by;
\begin{eqnarray}
w=\rho^{-3/2}\;\;\;\;\;\;\;\; D=(\frac{15}{64R_0})^{1/4}
\end{eqnarray}
so that the metric takes the form
\begin{eqnarray}
ds^2 = 4D^2 \rho^{-1} (-dt^2 + dz^2 + d\rho^2) + \frac{\rho^{-2}}{4D^2} d\phi^2,
\end{eqnarray}
or in a new coordinate system with $\bar\rho = 4D\rho^{1/2}$
\begin{eqnarray}\label{n5}
ds^2 = 64D^4 {\bar \rho}^{-2} (-dt^2 + dz^2) + d{\bar \rho}^2 + 64D^2 {\bar\rho^{-4}}d\phi^2.
\end{eqnarray}
Now if one takes $R_0 =4\Lambda$, as in the case of the Einstein field equations in
the presence of $\Lambda$, then the constant $D$ introduced above is related to the
cosmological constant through
\begin{eqnarray}\label{n3}
D^4 = \frac{15}{256}\Lambda^{-1}.
\end{eqnarray}
Of course one should be careful with this kind of identification of the spacetime
parameter as will be discussed later.\\
{\bf B})-The other solution with constant non-zero Ricci scalar could be found by
starting from the definition
$p:=p(w)=\frac{dw}{dr}$ in terms of
which the equation (19) can be integrated to obtain the following first
integral:
\begin{eqnarray}\label{01}
\ln(w)-\frac{1}{3}\ln(3p(w)^{2}-2p(w)c_{2}-c_{2}^2+4p(w)c_{1})\nonumber
\\-(c_{2}\frac{1}{3}\arctan(\frac{\frac{3}{2}p(w)-\frac{1}{2}c_{2}+c_{1}}
{\sqrt{-c_{2}^{2}+c_{2}c_{1}-c_{1}^2}})+\nonumber
\\c_{1}\frac{2}{3}\arctan(\frac{\frac{3}{2}p(w)-\frac{1}{2}c_{2}+c_{1}}
{\sqrt{-c_{2}^{2}+c_{2}c_{1}-c_{1}^2}}))(\sqrt{-c_{2}^{2}+c_{2}c_{1}-c_{1
}^2})^{-1}+C=0.
\end{eqnarray}
Also in terms of the same function the radial coordinate and the metric
functions are given by;
\begin{eqnarray}
r=\int \frac{dw}{p(w)}\\
u(w)=1/2\int \frac{p(w)+c_{2}}{wp(w)}dw\\
k(w)=\int \frac{p(w)+c_{1}}{wp(w)}dw.
\end{eqnarray}
Again it does not seem to be an easy task to find solutions for
the complicated  equation (\ref{01}), but in principle for each set of
the values for the integration constants $c_{1}$, $c_{2}$ and $C$
we have a solution for $p(w)$ and correspondingly solutions for
metric functions $k(r)$ and $u(r)$. One such solution, looking at
equation (\ref{01}), could be obtained by choosing $c_{2} = 2c_{1}$. In
this case, defining $A\equiv e^{C}$, we find the following
solution for the metric functions
\begin{eqnarray}\label{02}
k(w)=\ln  \left( w \right) -\frac{1}{\sqrt{3}}\,{\it arctanh} \left( {\frac
{\sqrt {{A}^{3}{w}^{3}+4\,{c_{{1}}}^{2}}}{2c_{{1}}}}
\right)
\end{eqnarray}
\begin{eqnarray}\label{03}
u(w)=\frac{1}{2}\,\ln  \left( w \right) -\frac{1}{\sqrt{3}}\,{\it arctanh} \left( {\frac
{\sqrt {{A}^{3}{w}^{3}+4\,{c_{{1}}}^{2}}}{2c_{{1}}}}
\right),
\end{eqnarray}
in which $w$ is a solution to the following equation
\begin{eqnarray}
3(\frac{dw}{dr})^2 -e^{3C} w^3 -4{c_1}^2 = 0.
\end{eqnarray}
Apart from the trivial solution $w={(-4{c_1}^2)}^{1/3} e^{-C}$
(which is indeed equal to zero by equation (19) for a constant $w$
and the fact that  $c_{2} = 2c_{1}$ ) one could show that there is
a solution in terms of the Weierstrass $\wp$ function
\footnote{Also known as Weierstrass elliptic function, is a doubly
periodic function. It is usually written either as $(z; g_1, g_2)$
or  $(z| \omega_1, \omega_2)$ in which $g_1$ and $g_2$ are called
elliptic invariants and are given in terms of the function's
semi-periods $\omega_1$ and $\omega_2$ \cite{11}.} as follows;
\begin{eqnarray}\label{04}
w =Weierstrass \wp(\frac{2^{1/3}}{6}\sqrt{3}e^C r + d, 0,
-4{c_1}^2) e^{-C} 2^{2/3}.
\end{eqnarray}
The constant curvature of the spacetimes given by the functions (\ref{02}), (\ref{03}) and (\ref{04}) could be
found through equation (7) in the coordinate system
($t,w,\phi ,z$), for which we find $R_0=A^3$. Now if again this is going to be
compared with the solutions
of the Einstein field equations in the presence of the cosmological
constant for
which $R_0=4\Lambda$, then the
correspondence $R_0 = e^{3C} = {4\Lambda}$ will fix the value of the
constant $C$  as
\begin{eqnarray}\label{05}
C = \frac{1}{3} ln 4\Lambda.
\end{eqnarray}
By looking at equation (5) it is clear that for constant Ricci scalar
solutions ($R = R_0$), no matter what the symmetry, $f(R)$ should
satisfy the relation $F (R_0) R_0 = 2f(R_0)$. Unlike the
solution with $R_0=0$, the above two solutions satisfy this relation for the commonly considered
model of  $f(R) = R- \frac{{\mu^4}}{R}$ with ${R_0}^2 = 3{\mu^4}$ if we have
\begin{eqnarray}\label{06}
D^8 = (\frac{5}{64})^2\frac{3}{\mu^4}\\
C = \frac{1}{6} ln (3{\mu^4})
\end{eqnarray}
in cases {\bf A} and {\bf B} respectively.
\section{discussion and summary}
Studies on the exact $f(R)$ gravity solutions are mostly restricted to
the spherically symmetric
case mainly due to the solvability of the equations and also more
importantly the fact that one
could compare the results with the solar system observations/experiments
based on the
schwarzschild solution as the spacetime metric around a spherical
mass such as Sun. Here we have examined cylindrically
symmetric solutions in metric $f(R)$ gravity in a general form given by (6).
Restricted by the complexity of the field equations we have only  examined
solutions with constant Ricci scalar. In the case of Ricci flat solutions we have found a generalized
form of a conical spacetime (with zero curvature) which, as a special
case, includes the cosmic
string spacetime. In the non-zero Ricci scalar case we have obtained
two new solutions and in both of them
a parameter is identified as the cosmological constant through the comparison of their Ricci scalar with
that of the modified Einstein field equations (in the presence of the cosmological constant).
Obviously neither of these spacetimes are asymptotically flat nor they behave regularly
in the limit $\rho \rightarrow 0$, and so one can not use their asymptotic behaviour
to compare their parameters with those solutions known in GR  having the same symmetry.
Therefore one should note that either of the identifications (\ref{n3}) and (\ref{05}),
is a very naive one in the sense that we have not
compared the corresponding  spacetimes (nor we have studied their
specifications) with one
already known in the context of Einstein GR. Even if there are any
solutions in GR comparable
to these solutions (which we are not aware of), since the field
equations in $f(R)$ gravity
are in general of higher order compared to their counterparts in GR, the
correct
identification should be made through matchings of different patches of
the whole
manifold. For example in the cylindrical case these might correspond to the exterior
and interior
solutions of a cylindrical shell if one of the parameters is going to
be interpreted
one way or another as the mass of the shell \cite{4}. \\
Another point need to be mentioned 
Incidentally it should be noted that the metric (\ref{n5}) has the same general form as the LT solution
introduced in \cite{6} (see also \cite{10}) \footnote{It is shown in \cite{6} that for a cosmic string in the
presence of a positive cosmological constant, one could write the metric in the following general form
$$ds^2 = cos^{4/3}(\lambda \rho)(dt^2 - dz^2) - d\rho^2 - \lambda^{-2}
\frac {sin^2(\lambda \rho)}{cos^{2/3}(\lambda \rho)} d\phi^2$$
in which $\lambda = \frac{1}{2}(3\Lambda)^{1/2}$ and other parameters are set equal to $1$ \cite{10}.}.
Unless one could find a solution to the equation (\ref{n4}) which exactly corresponds to the LT solution
in Einstein gravity, it seems a reasonable conjecture to say that the solution (\ref{n5}) is the $f(R)$
analogue of LT solution in GR. Finally it should be noted that in the vacuum case, for $R= constant$ the metric and the Palatini $f(R)$ formulations are dynamically equivalent \cite{Fera}. Therefore all the above solutions are also solutions in the Palatini gravity with the same $f(R)$ \footnote{We are grateful to M. L. Ruggiero for pointing this out to us.}.
\section{Acknowledgement}
The authors would like to thank University of Tehran for
supporting this project under the grants provided by the research
council. M. N-Z also thanks the University of Tehran's center of
excellence in the structure of matter for the partial support.

\end{document}